\documentstyle[11pt,epsf,rotate,fleqn]{article}
\newcommand{\Del}{$\Delta$}
\def\beq{\begin{equation}}
\def\eeq{\end{equation}}
\def\bea{\begin{eqnarray}}
\def\eea{\end{eqnarray}}
\def\eqref#1{Eq.~(\ref{eq:#1})}
\def\eqlab#1{\label{eq:#1}}

\def\figlab#1{\label{fig:#1}}

\def\NP#1#2#3{Nucl. Phys. {\bf #1} (#2) #3}
\def\PL#1#2#3{Phys.~Lett. {\bf #1} (#2) #3}
\def\PR#1#2#3{Phys.~Rev.~{\bf #1} (#2) #3}
\def\PRL#1#2#3{Phys.~Rev.~Lett. {\bf #1} (#2) #3}
\def\AP#1#2#3{Ann. of Phys. {\bf #1} (#2) #3}

\def\vslash#1{\mbox{/\llap #1}}    % p-slash

\begin{document}
\thispagestyle{empty}
\vspace{20pt}

\begin{center}
{\LARGE
Lepton-Pair Production in Virtual Compton Scattering off the Proton}\\
\vspace{20pt}
{\large  A. Yu. Korchin$^{a,b)}$, O. Scholten$^{a)}$ and F. de
Jong$^{c)}$}\\
\end{center}
\vspace{7pt}
{$^{a)}$Kernfysisch Versneller Instituut, 9747 AA Groningen, The
Netherlands\\[0.1cm]
$^{b)}$National Science Center ``Kharkov Institute of Physics and
Technology'', 310108 Kharkov, Ukraine\\[0.1cm]
$^{c)}$Instit\"{u}t f\"{u}r Theoretische Physik,
Universit\"{a}t Giessen,
35392 Giessen, Germany.
 %\\[0.1cm] \indent  $^1$Present address. }
\vspace{2cm}
 
%\DRAFT

\centerline{\bf Abstract}
\begin{quote}
  We show that lepton-pair production in Virtual Compton Scattering offers, 
through interference with the well-known Bethe-Heitler process, a sensitive 
probe to learn the longitudinal response of resonances and 
the electromagnetic nucleon form factors. This 
interference can be measured directly in terms of an asymmetry.  
The role of off-shell effects in the N-N-$\gamma$ vertices 
is investigated as well. An additional N-N-$\gamma$-$\gamma$ contact term in 
the amplitude, included to ensure gauge invariance of the model, cancels a 
substantial part of the off-shell effects.
\end{quote}

\bigskip
\noindent
{\bf 1996 PACS} numbers: 13.30-a, 13.40.-f, 13.40.Gp, 13.40.Hq, 13.60.Fz
 \Del-isobar.
\bigskip

Submitted to Phys. Lett. B, january 13, 1997

%\vfill
%\hfill  \today
\newpage

%\section{Introduction}

We investigate Virtual Compton Scattering (VCS) in the time-like region of
the photon momentum $k^2$, the
process where the incoming photon is real with a virtual outcoming photon, 
decaying into an electron-positron pair. One important reason for
investigating VCS is to extract the nucleon form factor in the `unphysical'
region, $k^2<4M^2$ ($M$ is the nucleon mass), where no data exist. 
In this letter physics is addressed related to
the more general electromagnetic structure of the nucleon, such as
off-shell effects in the N-N-$\gamma$ vertex and the longitudinal response
of nucleon resonances.  The main complication in VCS is the huge
`background' due to the well known Bethe-Heitler (BH) process. In
Ref.~\cite{Scha95} the electromagnetic vertex in the unphysical region has
also been considered where it was shown that, due to this large BH
contribution, only at large opening angles of the lepton pair
information on the time-like form factor could be extracted. As shown in
this letter the interference with the BH-background can be exploited to make 
VCS a sensitive measure for nucleon structure.

\begin{figure}[hbtp]
\begin{center}
\leavevmode
\epsfysize=14cm
\rotate[r]
{\epsfbox[114 96 344 800]{c: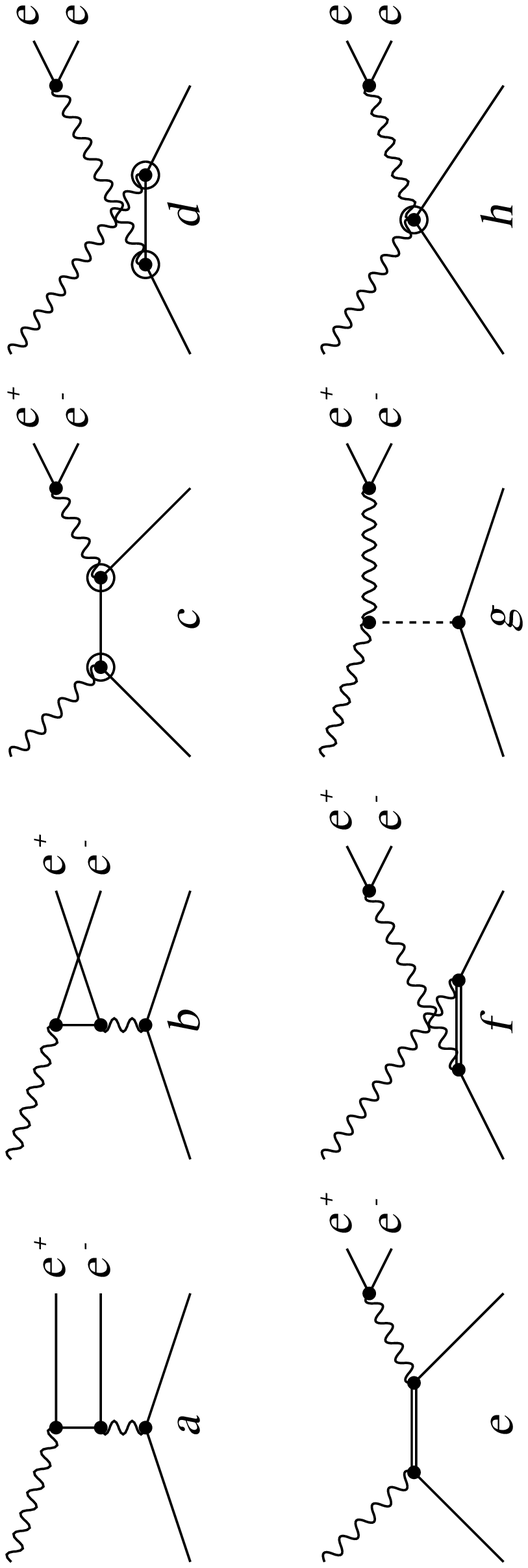}}
\caption[feyn]{\figlab{feyn}
The tree-level Feynman graphs included in the calculation where diagrams a and b
correspond to the Bethe-Heitler process. Circled dots represent vertices
where off-shell effects are considered, double lines represent nucleon
resonances, and the dashed line is a $\pi^0$.
}\end{center}
\end{figure}

Following an idea originally proposed in~\cite{Alv73} and more
recently investigated in~\cite{Lvo96} the BH-process can be used
to ones advantage. The lepton pair emerging from the BH process
interacts with two photons (see  Fig.~1a,b) and thus has positive
charge-conjugation parity (C-parity)~\cite{Bjo58}. The lepton pair from the nuclear
process (Fig.~1c-h) interacts with a single (virtual) photon and
thus has negative C-parity. The different symmetry of the two matrix
elements under C-parity implies that of
the three terms in the differential VCS cross section,
\beq%-------------------------------------------------------------------------
d\sigma(e^+,e^-) \propto \left| {\cal M}_{BH} + {\cal M}_N \right|^2= |{\cal
M}_{BH}|^2 + |{\cal M}_N|^2 + 2 Re ({\cal M}_{BH} {\cal M}_N^*) \;,
\eeq%=========================================================================
the first two terms on the right-hand side are symmetric under the
interchange of $e^{+}$ and $e^{-}$ while the last one is
anti-symmetric. The Bethe-Heitler-nuclear (BH-N) interference can thus be
measured directly through the asymmetry
\beq%-------------------------------------------------------------------------
A_{BH-N}={d\sigma(e^+,e^-) - d\sigma(e^-,e^+) \over d\sigma(e^+,e^-) +
d\sigma(e^-,e^+)} = {2 Re({\cal M}_{BH} {\cal M}_N^*) \over |{\cal
M}_{BH}|^2 + |{\cal M}_N|^2 } \;.
\eeq%=========================================================================
where $d\sigma(e^-,e^+)$ is the cross section under the same kinematic
conditions as $d\sigma(e^+,e^-)$ with only an interchange of the leptonic
charges. In the following the discussion is limited to this asymmetry.

To have a clear link to the nuclear process we have chosen as kinematical 
variables not the momenta of the electron ($k_1$) and positron ($k_2$), but 
instead $k^2$ (where $k=k_1 + k_2$, the momentum of the virtual photon in 
the diagrams in Fig.~1c-h), $\theta_k$ (the angle between $\vec{k}$ and the 
incoming real photon), $\theta_d$, (the polar angle between $\vec{k}$ and 
$\vec{k_d}=\vec{k_1}-\vec{k_2}$), and  $\phi_d$ (the azimuthal angle between 
the reaction plane and that of the $e^{+}e^{-}$-pair).  In these variables 
interchanging the leptonic charges corresponds to changing $\vec{k_d}$ to 
$-\vec{k_d}$. It should be noted that due to reflection symmetry with 
respect to the reaction plane the asymmetry vanishes for $\phi_d=90^o$, the 
results are henceforth quoted for the in-plane conditions $\phi_d=0^o$. For 
a polarized real photon or a polarized target this reflection symmetry is 
broken~\cite{Lvo96}.

In the calculations only the \Del- and the Roper-  (P$_{11}$) resonances 
have been included. The width of the resonances is generated in the 
unitarized K-matrix approach through the coupling to the one-pion decay 
channel. Parameters in the N-$\gamma$-resonance vertices have been choosen 
to obtain a best fit to the real Compton-scattering cross sections at photon 
energies $E_{\gamma}=150 \div 370$ MeV. For details of the model and the 
default parameters we refer to Ref.~\cite{Sch96} (parameter set\#2). 

In the present discussion of VCS we will focus on terms in the 
electromagnetic vertices that cannot be studied in real Compton scattering.  
Since their effects generally increase with increasing $k^2$, we limit the 
present discussion to $k^2$ close to the kinematical limit 
($\sqrt{s}=\sqrt{k^2}+M$). Even in these kinematical conditions the nuclear 
matrix element is still dominated (to more than 90\%) by  
transverse-polarized virtual photons, the longitudinal response plays a much 
smaller role than  expected. Inclusion of nucleon resonances in the 
calculation is thus important. Since only the lowest two nucleon resonances 
are included in the model we performed calculations at $E_{\gamma}=500$ MeV 
in the lab, corresponding to $\sqrt{s}=1.349$ GeV, and at $\sqrt{k^2}=406$ 
MeV (photon invariant mass for the nuclear contribution).  At this energy 
also higher resonances such as especially the D$_{13}$ may be important for 
the detailed results, but not for the general discussion and conclusions in 
this letter as we have verified. For the $k^2$ dependence of the nucleon and 
N-$\gamma$-resonance transition form factors we used the dipole fit for 
negative $k^2$ and Vector Meson Dominance (VMD) model for positive $k^2$.
Note that eventhough the longitudinal response contributes little to the nuclear matrix 
element, it strongly affects the asymmetry.

\begin{figure}[hbtp]
\begin{center}
\leavevmode
\epsfxsize=7cm
%\rotate[r]
{\epsfbox[36 50 338 774]{c: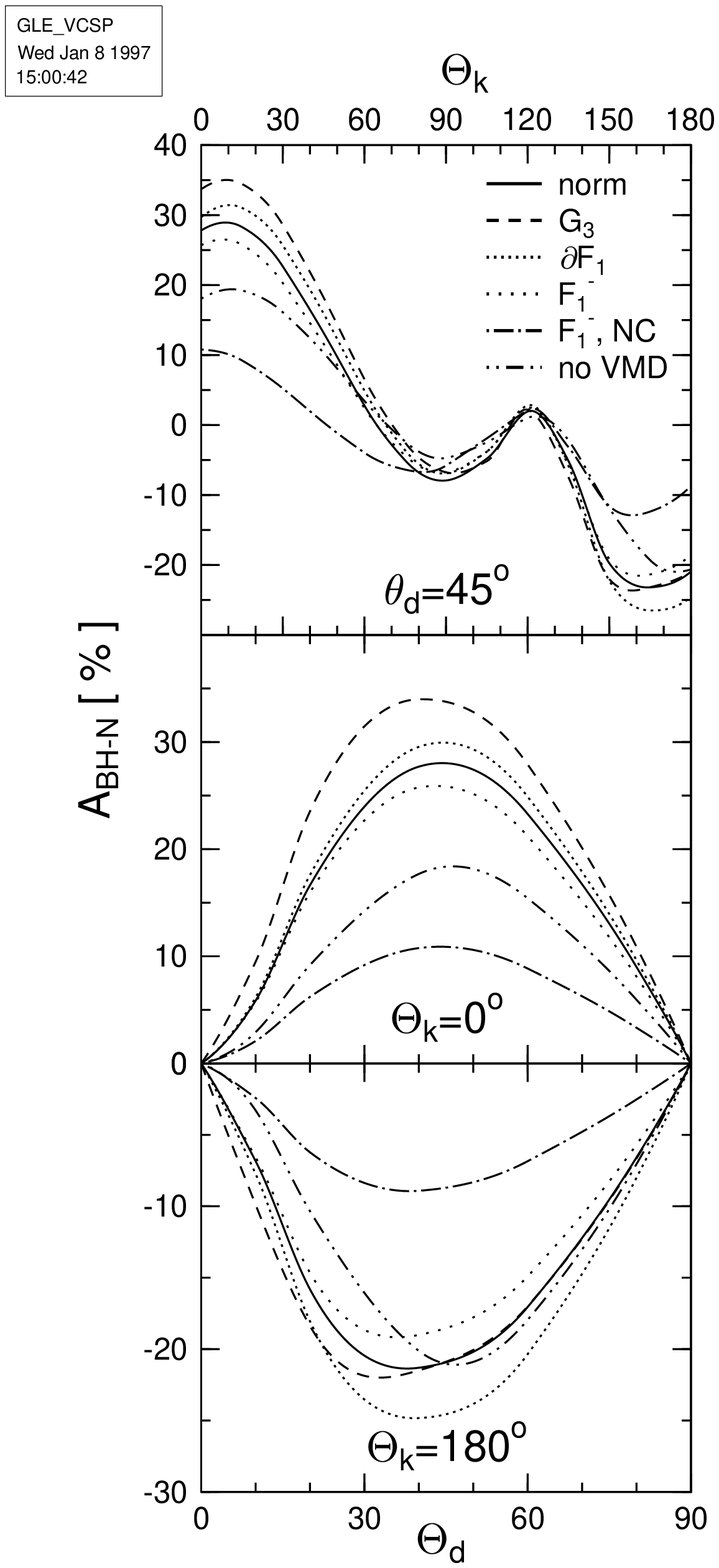}}
\caption[asym]{\figlab{asym}
Asymmetry in the center-off-mass system calculated as discussed in the text.
In the top panel the
opening angle of the lepton pair, $\theta_d$, is kept fixed, while
the direction $\theta_k$ of the lepton pair momentum varies.
In the lower two panels the asymmetry is plotted as function of $\theta_d$
at fixed $\theta_k$. The curves `G$_3$' show the
contribution of the longitudinal coupling to the \Del-resonance, the curves
`$\partial F_1$' and `$F_1^-$' display the effect of the inclusion of the
off-shell form factors in the N-N-$\gamma$ vertex.
The effect of switching off the contact term \eqref{contact} (for the calculation
`$F_1^-$') is shown by the dashed-dotted curves. The effect of the 
$q^2$-dependence in the form factors can be seen from the curves `no VMD'.
}\end{center}
\end{figure}

The VCS matrix element depends strongly on the parameters in the 
N-$\gamma$-resonance vertices. For simplicity and as an example we limit 
ourselves to the discussion of the longitudinal part of the N-\Del-$\gamma$ 
vertex~\cite{Jon73}, proportional to G$_3$, which does not contribute to 
real Compton scattering. 
In Fig.~2 the curves labelled `G$_3$' show the 
calculated asymmetry $A_{BH-N}$ for G$_3 =20$ which can be compared with the 
calculations using default parameters (G$_3=0$, curves labelled `norm'). At 
backward angles (lower panel of Fig.~2) the effect of G$_3$ is not just an 
overall enhancement of the interference pattern but introduces a change in 
the structure. This is due to a changing ratio of longitudinal v.s.\ 
transverse response which is reflected in a different 
dependence~\cite{Kor96} on the opening angle $\theta_d$. We also find that 
the asymmetry depends strongly on the off-shell parameter 
$z_3$~\cite{Pas95}. The results shown in Fig.~2 have been obtained for 
$z_3=-0.5$, at $z_3=0$ the dependence on G$_3$ almost vanishes.

It is known that the N-N-$\gamma$ vertex for off mass-shell nucleons has a
more complicated structure~\cite{Nau87,Tie90,Bos93,Doe95} than that for the free nucleon. The corresponding
form factors depend on all Lorentz invariants that can be
formed from the momenta at the vertices, not only the photon invariant
mass $q^2$. In particular we will consider the dependence on the
invariant mass $W$ of the intermediate nucleon in the diagrams Fig.~1c,d.
For sake of simplicity we limit ourselves to a half-off-shell vertex of the structure
%(where it is implied that either initial or final nucleon is off-mass-shell)
\bea%-------------------------------------------------------------------------
\Gamma^{\mu}_{NN\gamma} &=& \hat{e} \gamma^\mu +
\tilde{F}_1 (q^2) {\cal O}_1 (p') (q^{\mu} \vslash{q} - q^2 \gamma^\mu)
{\cal O}_1 (p) \nonumber \\
&&+ {F}_2 (q^2) {\cal O}_2 (p') i {\hat{\kappa} \over 2M} \sigma^{\mu \nu}
q_{\nu} {\cal O}_2 (p)
\eqlab{vertex}
\eea%=========================================================================
based on a Taylor-series expansion, with
\beq%-------------------------------------------------------------------------
{\cal O}_i (p)= 1 + (A_i + B_i {\vslash{p} \over M} ) ({\vslash{p} \over M} -1) \;,
\eqlab{off}
\eeq%=========================================================================
and $q=p'-p$ ($\hat{e}=1(0)$ for the proton (neutron)). Here $F_1 (q^2) =\hat{e} - q^2 \tilde{F}_1 (q^2)$
and $F_2 (q^2)$ correspond to the usual Dirac and Pauli form
factors in the VMD model. The off-shell dependence is
introduced by the operators ${\cal O}_i$ normalized such that
\beq%-------------------------------------------------------------------------
{\cal O}_i (p) u(p)=u(p) \;,
\eqlab{u}
\eeq%=========================================================================
 when $u(p)$ is a positive-energy solution of
the free Dirac equation. The off-shell effects are illustrated through the
relations
\bea%-------------------------------------------------------------------------
{\partial {\cal O}_i (p) \over \partial \vslash{p} } u(p)&=&{A_i +B_i
\over M} u(p)
\eqlab{der}\\
{\cal O}_i (p) v(p)&=&(1- 2 (A_i - B_i ))\, v(p) \;,
\eqlab{v}
\eea%=========================================================================
where $v(p)$ is a negative-energy state. Choosing $A_i=B_i=0$ corresponds
the conventional form for the free nucleon vertex.

The diagrams in Fig.~1c-g satisfy gauge invariance if off-shell
effects are absent. Inclusion of these leads to a violation of
current conservation and thus requires additional terms in the reaction
amplitude eventhough the vertices in \eqref{vertex} obey the Ward-Takahashi 
identity for the reducible vertex. Using constraints imposed by the current conservation
with respect to initial and final photons (see~\cite{Sche96}) one can
construct an effective N-N-$\gamma$-$\gamma$ vertex (Fig.~1h) to correct 
gauge invariance.
The corresponding vertex (called here the contact term) for the off-shell
dependence in \eqref{off} takes the form
\bea%-------------------------------------------------------------------------
\Gamma^{\mu\nu}_{NN\gamma\gamma^*}&=& 2\hat{e} \tilde{F}_1 (k^2)
{(A_1 - B_1) \over M} \left( k^2 g^{\mu\nu}-k^{\mu} k^{\nu} \right)
\nonumber \\
&& + 2 \hat{e} \tilde{F}_1 (k^2) {B_1 \over M^2} 
(p+p')^{\mu} (k^2 \gamma^{\nu} - k^{\nu} \vslash{k}) \nonumber \\
&& + \hat{e} {\hat{\kappa} \over 2M} {(A_2 - B_2) \over M} \left[
 (\gamma^{\mu} \vslash{q} \gamma^{\nu} - \gamma^{\nu} \vslash{q} \gamma^{\mu}) +
 F_2(k^2)( \gamma^{\mu} \vslash{k} \gamma^{\nu} 
         - \gamma^{\nu} \vslash{k} \gamma^{\mu} ) \right] \nonumber \\
&&  + 2 i \hat{e} {\hat{\kappa} \over 2M} {B_2 \over M^2} \left[
-(p+p')^{\nu} \sigma^{\mu\rho} q_{\rho} + F_2(k^2) (p+p')^{\mu}
\sigma^{\nu\rho} k_{\rho} \right] \; .
\eqlab{contact}
\eea%=========================================================================
Here the index $\mu$ ($\nu$) and momentum $q$ ($k$) refer
to the real (virtual) photon. One should keep in mind that the off-shell 
effects may be different for different representations of the same theory. 
This has been shown for the case of real Compton scattering off the 
pion~\cite{Sche95} and in a different context in~\cite{Dav96}. The off-shell 
effects addressed in the present paper are defined within a particular 
representation for the effective Lagrangian. Note that the structure of
this contact term is not unique and ambiguities in its structure
may be as important as the effect of the off-shell dependence itself.

 The physics of the contact term can be understood as follows. Introducing 
an explicit nucleon-momentum dependence in the vertex corresponds in 
coordinate space to creation and annihilation of the nucleon at different 
positions, i.e.\ the introduction of some general finite size effects. When 
the nucleon is charged this corresponds to violation of current 
conservation. An additional contact term should therefore be included 
corresponding to the coupling of the photon to the current which was not 
accounted for. Alternatively, in a one-pion loop model for form factors, 
off-shell effects arise through the fact that the (virtual) photon can 
couple to both the nucleon and the pion in the loop. Inserting this vertex 
in a calculation where one deals with two photons, like Compton scattering, 
one should in addition take into account that the other photon couples also 
to the charged particles in the loop. This effect leads to an effective 
N-N-$\gamma$-$\gamma$ vertex.

The VCS matrix element depends strongly on off-shell dependences in both
the Dirac and Pauli terms in \eqref{vertex}.
 Since the effect of the Pauli term can also be investigated in real
Compton scattering we will limit the present discussion to the Dirac term
and henceforth take $A_2=B_2=0$. One can
distinguish two cases:
\begin{itemize}
\item[$\partial F_1$:] $A_1=B_1$ in \eqref{vertex} corresponds to the equal coupling for negative-
and positive-energy states (see \eqref{v}) and a non-vanishing derivative of the
coupling at the on-shell point (see \eqref{der}). The calculation for
$A_1=B_1=1$ is labelled `$\partial F_1$' in Fig.~2.
\item[$F_1^-$:] $A_1=-B_1$ in \eqref{vertex}, in contrast, corresponds to a the vanishing
derivative and a different coupling for positive- and negative-energy
states (compare \eqref{u} and \eqref{v}). The calculation for $A_1=-1$,
$B_1=1$ is labelled `$F_1^-$' in Fig.~2.
\end{itemize}
%The order of magnitude for the parameters $A_1$ and $B_1$ is taken from the
%calculation of Ref.~\cite{Doe95}.
 From Fig.~2 it is apparent that both
dependences give rise to comparable effects in the asymmetry of about 5\% in 
$A_{BH-N}$. Under conditions further away from the kinematical limit,
which means smaller values of the photon invariant mass at fixed $\sqrt{s}$,
the effects of these terms becomes even less.

As remarked, the construction of the contact term \eqref{contact} is not
unique and we therefore compare results with a calculation in which the contact
term is switched off. This leads to a violation of current conservation,
however, it may give an estimate of the importance of this contact term. In the
`$\partial F_1$' calculation the contact term does not affect the results,
while in the `$F_1^-$' calculation it appears responsible for suppressing
most of the off-shell dependence (compare curves `$F_1^-$' and `$F_1^-,NC$' in
Fig.~2).

The calculations of ref.~\cite{Doe95}, based on a one-loop model and VMD, 
indicate that the values of the parameters $A_1$ and $B_1$ are of the order 
of 0.1, much smaller than the values of the off-shell parameters $A_2$ and 
$B_2$ that enter in the Pauli form factor which are of the order of $1/4$. 
As a result for realistic cases one expects the off-shell effects to be 
negligible in the asymmetry.

The $k^2$ dependence of the form factors is also reflected in the 
asymmetry.  Neglecting this dependence in all vertices changes the results 
considerably (see Fig.~2, curves labelled `no VMD'). At the backward angle the 
interference pattern is affected which offers the possibility to disentangle 
the $k^2$ dependence from off-shell effects. Therefore measurement of the 
asymmetry allows for a verification of the VMD model in the region $k^2 < 
M^2$, which could be complementary to the suggestion in~\cite{Scha95} to 
study form factors in the dilepton mass spectra under symmetrical conditions 
($\theta_k =0^o$ or $180^o$, $\theta_d =90^o$). However, to obtain 
information on the form factors in the interesting region $k^2 \approx 
m_{\rho}^2$ ($m_{\rho}$ is the mass of the $\rho$-meson) one has to study 
VCS at higher photon energies where, of course, contributions of the higher 
nucleon resonances should be accurately taken into account.

It can be concluded that the asymmetry in VCS offers an interesting tool to 
study the longitudinal response of nucleon resonances and electromagnetic 
form factors of the nucleon. In view of the results of~\cite{Alv73} we 
believe that in modern experiments it should be possible to distinguish  
the different effects. The most interesting region is at backward 
virtual-photon angles close to the kinematical limit where from the 
$\theta_d$ dependence it appears to be possible to verify the VMD model and 
to separate the longitudinal response of resonances.

\vspace{1cm}

We acknowledge discussions with S. Nagorny and A.E.L. Dieperink.
One of the authors (A.Yu.K.) acknowledges the financial support from
the Nederlandse Organisatie voor Wetenschappelijk Onderzoek (NWO).
He would also like to thank the staff
of the Kernfysisch Versneller Instituut in Groningen for the kind hospitality.

%\newpage

% \begin{itemize}
% \item[Fig.~1] {
% .}  \end{itemize}
\end{document}